# Improved Power System of the Future

Mario Rabinowitz
Armor Research, 715 Lakemead Way, Redwood City, CA 94062-3922
Mario715@earthlink.net

**Abstract**

This paper is intended to provide an insight into physics and engineering that can modernize electric power systems. Topics covered are Flexible ac transmission systems (FACTS), Custom Power, Greatly improved Capacitors, Electrical Insulation, Distribution Cables, Improved Polymeric Insulation, Underground Vault Explosions, Fault Location, Smart Cables, Neutral and Ground, Corrosion and Protection, Conventional Transformers, Compact Transformers, Ferroresonance, and Solid State Transformers.

## FACTS

### Just the plain FACTS

Flexible ac transmission systems (FACTS) is a new up-grade technology that can increase the use-capacity of existing transmission and distribution assets, as well as change the impedance of lines in affecting how power is routed along these lines. Although similar narrower concepts like the Static Var Compensator (SVC) were already in existence, the broader FACTS concept was developed by EPRI. FACTS was made possible by the advances in recent years of power electronics, utility-dedicated software and microcomputers, and fiber optic transmitters which permit signals and readings to be sent to or from high voltage levels. An essential ingredient in FACTS is the rapid and precise switching in and out of large capacitor banks. This is made possible by advances in solid state switches such as thyristors that makes them orders of magnitude faster, more precise, and more reliable than their mechanical switching counterparts, and also far superior to their gas discharge counterparts. Switching by thyristor-constituted systems can (within limits) control phase angle, impedance, voltage, and current in ways that would otherwise not be possible with mechanical breakers and switches. This is a major new development in the way utilities can increase line load so that the issue of new lines and new access laws need not be broached. For high power, the thyristor power loss in the forward direction is not negligible. However, wide-bandgap semiconductors may alleviate this. Scoping studies indicate the costs and cost savings achievable through FACTS. Two such studies are available as EPRI reports EL-6943 volumes 1 and 2; and another resource is the Proceedings of FACTS conferences, report TR100504.

Let us look at a few ways in which FACTS can operate. Switching capacitor banks in series into overhead delivery lines reduces the impedance of the lines. This process is intuitive as capacitive reactance simply subtracts from inductive reactance in a series circuit, decreasing the overall circuit impedance. An inductor can be switched in from the line to ground to suppress voltage surges. Less intuitive is the way a static var compensator operates to raise the voltage of an overhead line when drops in voltage threaten stability. Here a thyristor switches in a capacitor from the delivery line to ground. One might expect this shunt reactance to further reduce the line voltage. However, the proper application of a shunt capacitor in parallel with the line can increase the power factor and consequently increase the voltage (reduce the voltage drop between the source and the load). This is usually limited to about an average 20 % increase, though 50% may be possible in some cases. When delivery line voltages drop to more than 80% of normal, a new type of shunt using a gate turn-off (GTO) thyristor does much better. This system connects a dc capacitor that delivers accurately timed pulses of voltage to the delivery line to bring it back to its correct operating level.

FACTS can alter the impedance of a line and influence the routing of power within a system and between systems, thus giving practicality to the concept of a national grid. In this use, it will be necessary to have a coordinated operation of FACTS devices. As this application grows and becomes complex, the coordination will have to be so fast and precise that it will have to be computer controlled. Presently, power flow over dc lines is controlled by power electronics since both voltage and current are determined by the rectifier and inverter stations at the two ends of the line. When a dc line is embedded in an ac grid, the stability of the ac grid is enhanced by providing damping action at the converter controls. FACTS can also supply suitable damping to stabilize an ac grid. Harmonics that are deleterious both to customers and the utility's turbo-generation system can be significantly reduced by FACTS.

**FACTS operating systems**

Now, let's see what operating systems have employed FACTS. In 1991, AEP tested thyristor switching of a series capacitor bank on a 345 kV line in West Virginia. Subsequently, WAPA installed a similar system on a 230 kV line in Arizona which increased power transfer from 300 to 400 MW. The world's first large-scale FACTS was put into service on the BPA 500 kV, 2500 MW line in Oregon in 1993. It is a thyristor-controlled series capacitor (TCSC) system. In 1995, TVA incorporated GTO thyristor modules capable of controlling 100 MW, which can interrupt current at any point in the ac cycle. However, many utilities are taking a conservative approach in waiting for solid state devices to become more cost-competitive and proven in the field at high voltage and high power levels for long periods of time before committing to the large capital investment associated with FACTS.

**Custom Power**

The next 20 years will present both technical and business opportunity challenges to utilities in their distribution of electrical energy. The evolving nature of distribution system operations and end-use equipment performance presents a mixed bag of tasks that will need to be well-executed for utilities to succeed in the new highly competitive marketplace. Customers are demanding a more complex as well as more attractive set of products from utilities. Concurrently, the changing industry environment places the returns to shareholders at risk. Electrical utilities will be aided in meeting both their customer needs for custom power and their own need to be fiscally successful by appropriate software packages. Distribution business strategies are as much a part of custom power delivery as are the technical devices needed to produce and deliver custom power.

A growing number of loads are sensitive to customers' critical processes, which have costly consequences if disturbed by either poor power quality or (even worse) power interruption. It is estimated that power quality problems cost U.S. industry $26 billion per year (cf. to Section on Energy Storage, etc.). Exacerbating this problem are an increasing number of loads which produce interfering noise on the power system. Regulators have already radically changed the business environment for the sale of electricity. They may soon do so again by imposing penalties with respect to power quality and reliability.

The expanded use of power electronic controls and equipment (e. g., in customer automation equipment) brings both desirable and detrimental aspects with it. On the positive side, it can result in productivity improvement. However, it also brings with it large inductive overvoltages which can also cause power quality problems for both customers and utilities. Similarly, the high efficiency motors, adjustable speed drives , and high efficiency lighting with electronic ballasts that are gaining popularity also contribute to increased power quality problems.

In response to this, a mission was embarked upon to develop a new class of solid state devices which will be ready in the near future to enhance power system operation by ameliorating aberrations in power quality. Power electronic components are used to switch power on or off in a matter of microseconds. A single phase disturbance on a 3-phase system can be alleviated by quickly transferring power from the undisturbed phases to the perturbed phase. Power can also be transferred to energy storage apparatus for later use in case the system is upset. By themselves, static var generators may introduce harmonics into the system. Proper filtering can alleviate this, or a superconducting generator (cf. Section on ...Superconducting Generators) operated as a synchronous condenser would not introduce harmonics at all.

**Greatly improved capacitors**

New fast and high power solid state switching devices are at the heart of FACTS and make it possible. The workhorses of FACTS are the capacitors. The older kraft paper capacitors have been replaced with improved all-film cross-linked polypropylene capacitors. Because they represent a significant capital investment which also leverages real estate and expensive isolated high voltage platform cost due to their large size, some consideration should be given to developing greatly improved, lower cost, higher power density capacitors. Since capacitors can be traced back to the Leyden jar in 1745, one might think that they are so well developed by now that there is little room for improvement. Until recently, this view would have been correct. However, exciting new developments at the Los Alamos National Laboratory (LANL) hold out the promise that capacitively stored energy density can be increased by more than a factor of a 100. LANL has made small PZT (lead zirconium titanate) capacitors that store 36.7 $J/cm^3$. This is 183 times higher energy density than conventional capacitors which typically store about 0.2 $J/cm^3$. They found no degradation in this performance after subjecting a PZT capacitor to 100 million cycles at 1000 Hz.

It is likely that the exceptional performance of these small PZT capacitors will decrease as they are made larger and made to handle high voltages. Nevertheless, even if this factor of 183 decreases to just a factor of 10 increase in power density and a factor of 2 less cost relative to a comparable conventional high power capacitor, this may translate into a significant cost reduction for FACTS. A large savings may also be possible where ordinary power line series compensation capacitors are employed.

**Electrical Insulation**

A good insulator (dielectric) must meet two primary requirements: It must have an electrical resistivity and a dielectric strength sufficiently high for a given application. Secondary requirements relate to thermal and mechanical properties. Occasionally, tertiary requirements relating to dielectric loss and dielectric constant must also be observed. For power systems, it is essential that the needed properties not deteriorate in a given environment and desired lifetime. The delivery of electrical power depends critically upon the performance of electrical insulation.

Any insulator with an operational dielectric strength of 100 to 300 V/mil (40 to 120 kV/cm) may be considered good. Many insulators listed with a dielectric strength more than an order of magnitude higher than this, are ordinarily stressed at these lower values in practical operation for two reasons. One is that the dielectric strength tends to fall off with thickness (more so for some materials than others). The other is that imperfections in the form of voids, cracks, filamentary defects, and so forth occur whenever long lengths and thick sections of insulating materials are manufactured. Dielectric strength is a measure of the electric stress on or through a dielectric. It deteriorates with the ingress of water and with elevated temperature.

Whenever two dielectrics are in series in an electric field, the electric stress E is higher in the medium of lower dielectric constant $\kappa$ by the inverse ratio of dielectric constants: $E_2/E_1 = \kappa_1/\kappa_2$. Therefore, the electric field is higher in a defect, which generally has a dielectric constant of 1, by a factor of 2 to 3 than it is in the insulator. Thus the greatest stress is placed on the weakest link. Furthermore as is evident from Table 1, the dielectric strength is much less in the defect than in the insulator. The defects or voids fill up with a self-destructive plasma that increases their size and exacerbates the effect. Schemes such as oil impregnation are used when possible to fill the defects to increase both their dielectric constant and their dielectric strength.

**Table 1. Properties of Various Insulators**

| Insulator | Dielectric Strength ($10^3$V/mil) | Diel. Const. | Tensile strength ($10^3$lb/in$^2$) | Flammability |
|---|---|---|---|---|
| Lexan | 8 - 16 | 3 | 8 | self-extinguishing |
| Kaptan H | 3 - 8 | 3 | 22 | self-extinguishing |
| Kel-F | 2 - 6 | 2-3 | 5 | no |
| Mylar | 4 - 16 | 3 | 20 | yes |
| Parylene | 6 - 10 | 2-3 | 10 | yes |
| Polyethylene | 1 - 17 | 2 | 4-6 | yes |
| Teflon | 1 - 7 | 2 | 4-5 | very low |
| Air (1 atm. 1mm) | 0.1 | 1 | 0 | no |

Note: Flammability is not for arcing byproducts. $10^3$V/mil = 4 x $10^5$V/cm. $10^3$lb/in$^2$ = 6.9 MPa.

One theoretical limit of dielectric strength may be reached when the phonon generation rate due to electron acceleration and collisions exceeds the ability of the insulator to conduct heat away. The dielectric would then fail by disintegrating. Other mechanisms can lead to lower limits. However, because of defects, it is as if the whole

- 6 -burden is placed on the component least able to handle it, and this generally results in the lowest limit.  It will be difficult to achieve large (order of magnitude) gains in dielectric strength in the near future, because of the difficulty in making large insulating sections defect free.  Even if insulators could be made defect free initially, at ordinary temperatures defects will appear spontaneously.  Perfect crystals do not exist above cryogenic temperatures because the minimization of the free energy implies that a crystal must have entropy associated with lattice imperfections.  So at non-cryogenic temperatures, defects will be generated to minimize the free energy.  Artificial gains can be achieved by reducing the margin of safety in operating closer to the dielectric strength limit, using thinner insulation, and compromising the longevity of the insulator.

**Distribution Cables**
**Extruded cables**

Extruded dielectric cables using aluminum or copper conductors are the mainstay of distribution systems. They have the advantage of eliminating the need for insulating fluids like oil which have the potential for unfavorable impact when they leak into the environment.  There was a trend for them to come into more extensive use in transmission systems with the development of such cables at 138 and 230 kV, until the relatively recent introduction of paper-polypropylene-paper oil static cables for use in the 115 to 345 kV range.  Research on extruded dielectric cables was directed at increasing their ratings to 345 kV, because compared to low-pressure oil-filled (LPOF) and high-pressure oil-filled (HPOF) cables using oil-impregnated cellulose tape insulation, extruded cables offer lower cost and are easier to install and maintain.  However, difficulties in making splices has curtailed their introduction at the higher voltage levels, but this has not been a problem at the lower distribution voltages.  Properly manufactured extruded insulation has advantages for use as the dielectric in low temperature cables since this avoids intensification of the electric field in the lower dielectric constant cryogen that is present in lapped tape insulation. [1, 2]

Of the three dielectrics that are commonly used, cross-linked polyethylene (XLPE) has a higher dielectric strength than ethylene propylene rubber, and XLPE is preferred over polyethylene because of polyethylene's lower softening point. Ethylene propylene rubber is sometimes preferred because of its greater flexibility in colder climates.   Water treeing in particular, and electrical treeing in general is still a problem.  New materials retardant to water treeing are being studied, as well as watertight cable design.  Cables impregnated with silicone or other additives seem to better withstand treeing.  One project is studying the addition of pressurized $SF_6$ to fill cables that are already full of

-6-

trees, and thus extend their useful lifetimes. (Interestingly when pressurized, the breakdown characteristics of these cables are more like a gas than a solid indicating that this may be a fruitful approach.) There is an incipient problem with underground vault explosions whose source is presently unknown. One possible solution is to develop insulations of generically different types as discussed in the next section.

**Improved Polymeric Insulation**

Significant introduction of new materials into power cables has not taken place since the development of polyethylene and polypropylene/paper cables. Is this because, by serendipity, we found the best dielectrics a quarter of a century ago and can hardly improve on them? This is not a derisive comment, as this often happens, as with barium titanate with a superhigh dielectric constant and remarkable ferroelectric properties. Or more likely, is it because it does not pay for a manufacturer to develop and introduce a new product into a stable market that will compete with his already successful product?

Improved insulation with higher dielectric strength would lead to thinner walled cables with a twofold benefit. The thinner wall would permit retrofitting existing cable ducts to hold higher ampacity cables as the conductor diameter could then be increased. Additionally, the thinner insulation would make the cables more flexible and easier to pull through the ducts. However, there is also a down side to thinner walls. The higher dielectric strength permits the same voltage drop across a thinner wall, but the concomitantly higher electric field may give rise to faster growth of deleterious water and electrical trees unless the material itself is somehow more resistant to such growth.

Possible polymers that look promising as extruded cable are TPX (4-methyllpentene-1), Aurem (Polyimide), and SPS (Syndiotactic Polystyrene). An alternative approach uses these and non-thermoplastic polymers in a laminar film form. PQ-100 (Polyquinoline) or Isaryl 25 can also be used in laminar form. To improve the dielectric properties, cross-linkable silicone or 1, 4-polybutadiene based resin can be used to permeate the cable and fill the voids. All these materials should be tested for electrical, mechanical, and combustion properties, as well as manufacturability. After narrowing the list to the most promising contestants, accelerated aging tests need to be conducted to determine their dielectric longevity. Finally the manufactured cost of the material will likely have to be less than $10/lb with a resulting distribution cable capital cost as low as $1 to $2/foot ($5000 to $10,000/mi). . While the dielectric properties of the extant materials will probably not improve a great deal, their manufactured cost may drop. We are always shooting at a moving target as illustrated in Table 2.

**Table 2. Target operating properties**

- Dielectric Strength            > 6000 V/mil = $24 \times 10^5$ V/cm.
- Dielectric Constant            < 3.5
- Loss tangent                    < $10^{-3}$
- Operating Temperature      > 130 °C

The dielectric constant needs to be low for 3 reasons. One is to minimize the electric field amplification in voids. Another is to keep the dielectric power loss low, which is proportional to the product of dielectric constant and loss tangent. The third is to keep the cable capacitance low, and minimize wasteful charging currents. Although it is difficult to compare dielectric data from different sources possibly using different test conditions, meeting these 4 target operating properties in a new material could lead to a 25% increase in ampacity.

**Underground vault explosions**

It is crucially important to find and correct the cause of underground vault explosions where access is provided for the distribution and transmission cables. These occur often enough to be a serious problem, but are not yet pandemic. This needs to be ameliorated before it becomes widespread. It is known that in the presence of arcing, large hydrocarbons break down into small hydocarbons, and that small hydrocarbons are explosive. Possibly due to bad splices, arcing of insulations like polyethelene and polypropylene may produce acetylene ($C_2H_2$), methane ($CH_4$), ethylene ($C_2H_4$), propylene ($C_3H_6$), etc. Such gases can cause explosions in confined areas. If this turns out to be the cause, this will greatly curtail underground power delivery of the near future until splices are sufficiently improved, or proper dielectrics for splices that are free of hydrocarbons are developed; or in the worst case scenario, a new insulation for the entire cable may have to be developed that is hydrocarbon-free such as teflon [carbon and fluorine], Kapton [carbon and nitrogen], Kel-F [carbon, fluorine, and chlorine], etc. (cf. Table 1).

**Attacking the underground explosion problem**

Unexpected and unknown explosive, combustible, toxic and other hazardous gases and particles can be introduced or released from a number of disparate sources in electric power delivery applications. It has been very difficult if not virtually impossible to determine which gases and/or particles caused a fire or explosion after the event, as the gases have been ignited and dispersed.  For example, in underground distribution networks, violent explosions have occurred in access areas, splice boxes, and underground vaults under manhole covers.[3. 4]  In one unfortunate instance, a woman was killed who was driving near the scene of an explosion.

Explosive, combustible, and other hazardous liquids, gases and particles can be released in a variety of ways.  These gases are sometimes generated by electrical arcing decomposition, hydrolysis, or thermal decomposition of materials.  These explosions can be prevented by precluding the precombustion culprit gases from building up and reaching a critical level as well as understanding their cause.  The underground explosion problem is being attacked in the following ways:

 1.  Prevention of the buildup of vapors and particles so that the concentration stays below the threshold for explosion.
 2.  Determination if the explosion was accidental, from natural sources, or due to man-made explosives e.g. by terrorists,  .
 3. Determination of culprit gases, preferably before an explosion, but also after an explosion.  If the culprit gases could be identified, this would be the first step in the avoidance of such onerous occurrences .

Until these approaches are adequately developed, methods of buttressing the vaults, etc. to lessen the effects of such explosions are being implemented.

**Fault location**

Fault location in complex urban distribution systems is presently a difficult, time-consuming, labor-intensive activity.  One of the widely used methods involves a capacitor discharge through the cables with utility personnel periodically positioned with headphone sensors listening for the noise associated with arcing in the cable.  This is known as the Thumper/Acoustic method.  In order to be able to detect the acoustic signal from an arc, the applied voltage and current are so large that they may cause additional damage to the cable.  Another method is the Earth Gradient method. A voltage source is applied across a cable, and an attempt is made to detect the leakage current to ground at the fault location by measuring voltage gradient differences in the earth.  This method has locale limitations, as well as problems related to spurious sources of voltage gradient.

A Fast Fault Finder is expectedto identify failed cable sections on underground distribution circuits with no taps, so they can be quickly isolated.  It uses transients created by the fault itself to locate the fault electronically.  Possibly by the judicious addition of high- frequency surge filters at junctures during cable installation, the Fast Fault Finder may be able to operate on even fairly complex circuits. If this is successful, because of its relatively low cost, it will probably be the method of choice for finding the location of faults in the future. For systems with wyes and tees, including networks, new methods are being developed for fault location involving very precise timing of electronic signals.

The high impedance fault detector is commercially available from GE, and will detect dangerous faults that were previously not possible to detect. It is mainly installed at substations, and can be used for both underground and overhead lines. Building on this technology, an overcurrent, ground, and adaptive relay function will be added, as well as low current, low voltage fault detection.

**Smart Cables**

Transmission and distribution (T & D) cables ultimately fail for various reasons including deterioration of insulation, overcurrents etc. Fault locators will help in detecting the location of a fault in the cable network after it has occurred in the sense of being a fully mature fault. It would be desirable, though more difficult, to determine where the fault will occur before it fully matures so that preventative action can be taken.

Properly designed smart cables with built-in microscopic sensors should be able to permit locating an incipient fault before it happens, and even more easily locating the fault after it has occurred. There may be enough changes in physical parameters to permit electronic determination that there is an incipient fault, and to locate it precisely. This would permit repair during routine maintenance. Although technical feasibility needs to be established, economic feasibility will determine whether smart cables become part of the power delivery system of the future. This will entail determining the economic value of taking remedial action before a fault has occurred, and thus avoiding the costs associated with down-time, locating the fault, and after the fault repair or replacement of the cable. It may well be that smart cables will find a niche in those circuits where power quality is at a premium.

**Neutral and Ground, Corrosion and Protection**

Previously, utilities' electrical copper neutral and ground wires were usually connected to and thus protected by someone else's buried steel piping, etc. This was no cause for concern in the past, as the corrosion of the steel was so slow that it could be ignored and occurred at someone else's expense.

However now that such steel grounding is not readily available, the utility's own electrical copper is vulnerable to corrosion because of ac voltages on the neutral, and soil variations. The electrolysis of the copper can be avoided by preventing transport of copper ions, and/or by preventing the exposure of the copper and formation of copper ions.

Without cathodic protection, copper corrosion may result in electrical shock hazards to utility personnel as well as the public; and serious property damage including loss of costly electrical equipment. In the future, increased public awareness and litigation will accelerate the need for mitigation measures. Because of the increasing trend toward undergrounding distribution cables, cathodic protection will be a requirement rather than an option in the future.

**Transformers**
**Conventional transformers**

Conventional power transformers have efficiencies that can exceed 98%, and generally have a track record of high reliability. While 98% efficiency is good, the 2% loss is primarily associated with core excitation losses which are present whether the transformer is loaded or not. The amorphous core steel transformer is an improvement on conventional transformers, which reduces no-load losses on distribution transformers by 60-70%, i.e. the efficiency can be increased to over 99% with the 24-hour/day core losses being cut more than a factor of 2.

There have been problems with large high-voltage transformers that require flowing oil to provide insulation and adequate cooling. In these transformers, oil flows through the windings to enhance heat transfer and at the same time assist the integrity of the electrical insulation. The oil flow transports electrical charge analogously to the operation of a Van de Graaff generator in which charge is carried by an insulating belt from ground to a high voltage terminal. In the transformer, accumulation of charge builds up the voltage which can exceed the dielectric strength of the insulation. The ensuing arc can lead to catastrophic failure of the transformer.

Two techniques can ameliorate this so-called static electrification problem. One is to increase the conductivity of the oil so that the charge leakage rate is high enough to prevent charge buildup. The limitation of this is that the dielectric strength of the oil is concomitantly reduced. The other technique is to place an additive such as 1,2,3 Benzotriazole (BTA) in the oil which reduces the electrification buildup of charge density . However, the BTA also decreases the oil conductivity thus decreasing charge leakage. New developments using combinations of additives show great promise in solving the transformer oil flow electrification problem. Thus less charge is removed at the oil-solid interface, and sufficiently high dielectric strength is maintained.

Solid encasement of transformer windings has always had the appeal of replacing flammable, explosive, and possibly toxic liquids like PCB. PCB (polychlorinated biphenyl) was classified as a highly toxic, non-biodegradable pollutant under Title 40 of the federal regulations code. Presently, alternate liquids are more costly and more flammable than PCB. This makes it all the more desirable to look at the possibilities for solid fillers. Solid encasement has had limited success. However, it encounters a twofold problem since it must serve the dual purpose of providing both insulation and cooling. From an insulation point of view, it must be void free for the same reason as discussed for cable insulation because the electric field is enhanced inside voids by the inverse ratio of dielectric constants. From a cooling point of view, thermal conductivity is much less for insulators than for conductors because phonon transport of heat is much less than electron transport at ordinary temperatures. However at cryogenic temperatures, phonon thermal conductivity in dielectrics can exceed that due to electrons and be adequately high. Hence for a cryogenic transformer, one may well reconsider the use of solid encasement. There is an interesting possibility that the Peltier effect can be incorporated for cooling in solid encasement, and other transformer insulation systems.

**Compact transformers**

A clear need and incentive exists to increase the power density in both transmission and distribution level transformers. The former are so large that cooling is a substantial problem. At the distribution level, there is a need to carry more power, and yet still fit the transformers in the existing limited underground vault space. The turns ratio of the primary and secondary windings establishes the voltage step-up or step-down characteristic of the power transformer. A single winding is sometimes used in a configuration referred to as an auto-transformer. The voltage ratio can be varied by a sliding contactor making electro-mechanical contact with the single winding. The fundamental principle upon which all transformers are based is that the induced voltage, V, equals the time rate of change of flux, $d\phi/dt$. Or equivalently, the induced voltage equals the inductance times the time rate of change of current, $L\, dI/dt$.

Two basic engineering concepts are immediately apparent from this. One is that at higher frequencies the power density goes up (but not without disadvantage). Thus transformers can be smaller since one can operate with less flux to get the same $d\phi/dt$. An example of this is the choice of 400 Hz frequency in airliners. However, this has its drawbacks as increasing the frequency increases the transformer impedance and losses (especially the core losses and leakage flux), inadvertently decreasing the power transmitted. The transformer reactance and hence the reactance voltage drop is directly proportional to the frequency. The hysteresis loss in the core is proportional to the frequency, and the core eddy current loss is proportional to the square of the frequency.

The other engineering concept, whose pursuit may be more fruitful, is that the power density can be increased by increasing the flux density. In a conventional transformer, laminated iron is used to concentrate the flux by lowering the reluctance of the magnetic circuit. However the maximum flux density, B, is limited to between 10,000 and 20,000 Gauss (1 to 2 Tesla) due to the saturation field of iron. The saturation field, $B_{sat}$, is where the permeability of the core drops precipitously and the core can no longer concentrate the flux. To reduce the size and cost of a transformer, transformers are designed to have the peak field not too far below the saturation field of the core so that the required flux can be retained by a minimum of core material. As we shall next see, this can be a problem.

The sun emits ionized particles into space on both a steady and a transient basis by means of what is called the solar wind. The steady interaction of the solar wind with the earth's ionosphere and geomagnetic field has no harmful effect on electric power networks. A much stronger and potentially adverse transient condition results from sunspot activity when the blast wave from a solar flare hits the earth's magnetosphere. Such geomagnetic storms induce offset current components in long east-west lines at northern latitudes, which can lead to core saturation and transformer damage as $B_{sat}$ is exceeded. Such electrical system problems occur about every eleven years during the peak of the sunspot cycle. The effects are exacerbated if the ground is poorly conducting in the vicinity of the network, causing more current to flow through the grounded conductors. To minimize these adverse geomagnetic effects, it is preferable not to have the peak field too close to $B_{sat}$ of the transformer core, which is antithetical to compaction i.e. making a transformer more compact.

Cryogenics presents one interesting route to compaction. Iron increases its saturation field from 2 T at 300 K to only 2.2 T at 0 K. Iron-cobalt ($Fe_{0.65}Co_{0.35}$), which is a strongly ferromagnetic material, increases its $B_{sat}$ to only 2.5 T at 0 K. Fortunately there is not a correlation between the Curie temperature, $T_{Curie}$, at which a material stops being ferromagnetic, and $B_{sat}$. Iron and $Fe_{0.65}Co_{0.35}$ have relatively high Curie temperatures of 1043 K and 1240 K. As Table 3 illustrates, materials with considerably lower $T_{Curie}$ can have considerably higher $B_{sat}$.

Using dysprosium, Dy, at ~ 77 K as the core material, together with cryoresistive Cu, Al, or perhaps even hyperconducting Be as the conductor, should be considered. The potential for achieving approximately a factor of 2 size reduction and higher power density in the conductor, as well as possibly even lower losses is worth looking into.

Some of the material costs look prohibitively high at present. What is not known is whether these costs can be substantially reduced if the market for these materials becomes considerably larger.

**Table 3. Ferromagnetic Materials**

| Material | $B_{sat}$, Tesla (at 0 K) | $T_{Curie}$, K |
|---|---|---|
| Dy | 3.8 | 87 |
| Ho | 3.7 | 20 |
| Er | 3.4 | 20 |
| Tb | 3.3 | 219 |
| Gd | 2.8 | 286 - 293 |
| $Fe_{0.65}Co_{0.35}$ | 2.5 | 1240 |
| Fe | 2.2 | 1043 |
| Co | 1.8 | 1390 |
| Ni | 0.64 | 630 |

The potential for a superconducting transformer at 77 K does not look nearly as great. The reason for this is that the large ac power losses in large fields coupled with a refrigeration inefficiency of 8 to 10 W/W appear prohibitive. The core could be eliminated with very large fields which exceed the saturation field of any known material. In principle, this could be achieved with a superconducting, hyperconducting, or possibly cryoresistive material. However, such an air core transformer would have a very large fringing field. This would likely not be acceptable for many practical reasons such as inducing losses in nearby conducting materials as well as the concern for biological effects of electromagnetic fields. Containing the field with a high permeability casing at sufficiently large distances that $B < B_{sat}$ is antithetical to size and cost reduction.

**Ferroresonance**

Ferroresonance is a complex electrical problem that is generally encountered only in distribution transformers because of single phase operation to which the large transmission level transformers are not subjected. Modern low-loss transformers are much more susceptible to ferroresonance. Ferroresonance is related to the fact that the resonance voltage across an inductor and capacitor in a series RLC circuit can be greater than the source voltage. Overvoltages due to ferroresonance run as high as 125 to 146% of the rated value. Such overvoltages can cause failure of the transformers themselves, auxiliary equipment, and even customer appliances and electronic equipment. Ferroresonance can lead to more than one response for the same switching conditions in a given circuit, much like chaotic phenomena.  The circuit may jump erratically from one non-sinusoidal mode to another.

One way of averting  ferroresonance is to use only 3-pole switches to connect or disconnect the distribution transformer from the circuit.  Another way is to use only triplex transformers with grounded-wye primary windings.  A third way is to use 3 single phase transformers with grounded-wye primary windings.  Another option that may be realized in the power delivery system of the future is to utilize solid state distribution transformers as discussed in the next section.

**10.4  Solid State Transformer (SST)**

In 1980 Bowers et al [5] described a solid state transformer (SST) with step-down and step-up capabilities. It used pulse width modulation to achieve ac voltage regulation without the need for large inductors.  At present, below kV voltages, solid state transformers based on  bi-directional switchmode technology (using the L dI/dt principle discussed earlier) efficiently transform voltages, and can also perform voltage and current waveshaping.  They can be more compact, have improved thermal management, and be insensitive to dc current components such as the troublesome ones from geomagnetic storms.  However at higher voltages, the power losses are presently prohibitively high by roughly an order of magnitude, making them impractical at this time for most distribution and transmission applications.  The higher the voltage, the greater the power loss.  There does not seem to be any reason in principle that further developments can't ameliorate the excessive voltage drop in the forward direction that now is encountered when the solid state switches have to withstand greater than kV voltages in the back direction.

The solid state transformer can be viewed as an outgrowth of the ac chopper voltage converter in which segments of the 60 Hz sinusoidal wave were chopped out so that each 8.33 msec half-wave was broken up into several remaining segments of a little over a msec each in duration.  This switching wave form resulted in low frequency voltage harmonics.  Presently, the quantum series resonant converter (QSRC) mode of operation replaces each 8.33 msec half-wave with about a dozen high frequency half-waves whose amplitudes model the amplitude of the 60 Hz wave.  The envelope of these ~ 720 Hz waves exceeds the 60 Hz wave a little so that the power delivered is comparable.  The QSRC eliminates low frequency harmonics.

An ideal bi-directional switching system may allow the SST to approach 95% efficiency, independent of load conditions.  If the presently large power loss at high voltage could be reduced, the SST may offer some attractive advantages.  Its initial-cost economic viability cannot be judged because solid state transformers are not being mass produced on a large enough scale as yet. Furthermore, its reliability and performance during highly stressful transient conditions such as fault currents, short circuits, and overloads have not been sufficiently determined. Although the early 1980's work originated in the U.S., most of the present work is being done outside the U.S.